\documentclass[sn-mathphys]{sn-jnl}

\jyear{2021}%

\theoremstyle{thmstyleone}%
\theoremstyle{thmstyletwo}%

\theoremstyle{thmstylethree}%

\usepackage{graphicx}
\usepackage{textcomp}
\usepackage{xcolor}
\usepackage{tabularx,adjustbox,booktabs}
\usepackage{colortbl}
\usepackage{multirow}
\usepackage{caption}
\usepackage{subcaption}
\usepackage{mathtools}

\begin{document}

\title[Employing Crowdsourcing in Higher Education]{Employing Crowdsourcing for Enriching a Music Knowledge Base in Higher Education}

\author*[]{\fnm{Vassilis} \sur{Lyberatos}*}
\email{vaslyb@ails.ece.ntua.gr}
\author*[]{\fnm{Spyridon} \sur{Kantarelis}*} 
\email{spyroskanta@ails.ece.ntua.gr}
\author[]{\fnm{Eirini} \sur{Kaldeli}}
\author[]{\fnm{Spyros} \sur{Bekiaris}}
\author[]{\fnm{Panagiotis} \sur{Tzortzis}}
\author[]{\fnm{Orfeas} \sur{Menis - Mastromichalakis}}
\author[]{\fnm{Giorgos} \sur{Stamou}}




\affil[]{\orgdiv{School of Electrical and Computer Engineering}, \orgname{National Technical University of Athens}, \orgaddress{\street{Zografou 15773}, \city{Athens},  \country{Greece}}}





\abstract{
   This paper describes the methodology followed and the lessons learned from employing crowdsourcing techniques as part of a homework assignment involving higher education students of computer science. Making use of a platform that supports crowdsourcing in the cultural heritage domain students were solicited to enrich the metadata associated with a selection of music tracks. The results of the campaign were further analyzed and exploited by students through the use of semantic web technologies. In total, 98 students participated in the campaign, contributing more than 6400 annotations concerning 854 tracks. The process also led to the creation of an openly available annotated dataset, which can be useful for machine learning models for music tagging. The campaign's results and the comments gathered through an online survey enable us to draw some useful insights about the benefits and challenges of integrating crowdsourcing into computer science curricula and how this can enhance students' engagement in the learning process. 
}




\maketitle

\section{Introduction}\label{sec1}

In recent years, we are witnessing an increasing number of studies that explore the use of crowdsourcing in education, embracing various disciplines, topics, and educational levels~\cite{jiang2018review}. Most of the existing work looks into crowdsourcing as a means of creating or assessing educational material and collecting feedback from students~\cite{alenezi,microlearning,wang2016employing}. Fewer studies tap into how involvement in a crowdsourcing experiment can serve an educational goal in itself~\cite{educsci12030151,khan2020completing}. We argue that incorporating crowdsourcing in the form of real-world and discipline-appropriate exercises in education curricula can bring about multiple benefits for students. First, the participatory nature of crowdsourcing can stimulate learning and engagement~\cite{psycology} and add a collaborative and creative touch to more traditional teaching procedures. Additionally, students can learn about the potential of crowdsourcing and how it can be useful within the context of their discipline.
Lastly, involvement in appropriate crowdsourcing tasks can help students familiarize themselves with important domain-specific concepts in a hands-on way and understand how the data collected through the process can be further exploited and enable new possibilities.

In tandem with its educational potential, crowdsourcing has been extensively used and studied in the context of citizen science, by providing the methodology and tools that enable the engagement of individuals who voluntarily contribute to knowledge production with a scholarly focus~\cite{shanley}.
Higher Education Institutions (HEI) can play a significant role in this context, by providing technical, material, and human resources, reinforcing open science policies, stimulating cross-disciplinary collaborations, and honing the competencies of new generations of scientists, researchers, and innovators as already mentioned. In such a setting, crowdsourcing can be seen as a driver of both citizen-enhanced open science and educational learning. Similarly, students are invited to play a dual role: act as citizens/contributors and as scientists/researchers.

The role that crowdsourcing can play in the Computer Science (CS) programs of HEIs remains relatively unexplored. Despite its close interrelations with IT practice and research, considering both the use of digital technology as a facilitator of crowdsourcing and, vice versa, the extensive application of crowdsourcing techniques in the IT domain, crowdsourcing is not among the subjects which usually make up a CS curriculum. In fact, crowdsourcing as a practice and technology remains peripheral to CS-relevant HE programs, with references to it and hands-on experiments being incidental (note that the crowdsourcing initiatives in education reviewed in~\cite{jiang2018review} include only one case study from CS).
Our main objective in this paper is to alleviate this gap by reporting on a case study that investigates how crowdsourcing can be incorporated into a CS curriculum as a component and facilitator of a mini-project assignment that can teach students useful lessons. As part of the case study, students were invited to participate in an online campaign with the aim to enrich the metadata of a music tracks collection. Students were then instructed to analyze the enriched dataset and apply semantic web technologies to construct a knowledge base and use it to extract useful information from it. 

In this context, our research set out to explore the impact crowdsourcing had along two main perspectives: learning outcomes achieved in terms of new knowledge and skills acquired within the scope and objectives of a specific CS course 
and how the participation experience was perceived by students
Concerning the educational gains and given the scope of the specific CS course we considered, we were particularly concerned with the extent to which crowdsourcing assists students to gain deeper insights into the structure and shortcomings of data, into the processes and technological infrastructures that can be used to acquire richer and higher-quality data, as well as into how the enhanced data can be further utilized.
Regarding the extent to which students felt engaged, we were mainly interested in affective characteristics, relating to feelings and attitudes.
Another direction we explored regards the role of technological tools, considering the requirements for certain features as well as the way in which the digital platform that was used influenced the experience of the participants. By putting themselves in the position of the contributor and platform end-user and by drawing on their capacity as CS students and technical experts, participants were able to provide insightful perspectives along this strand of inquiry. 

Overall, the case study exemplifies how crowdsourcing can fit into a CS course and serve its intended didactic objectives. By describing the methodology that was followed, from the data curation and the campaign setup to the exploitation of results and the evaluation approach, the technological tools that were used, the challenges that were encountered, and the way in which these were overcome, the current paper points both to the benefits as well as the limitations of the approach and can thus serve as a paradigm and as a source of inspiration for incorporating crowdsourcing in CS curricula and beyond.
Besides the educational benefits, in line with the principles of open and citizen science, the data generated by the case study is further processed and made openly available, thus contributing to ongoing developments in research on music tagging research.

\section{Related Work}

According to the typology suggested in \cite{jiang2018review}, crowdsourcing is used in educational activities to serve four main objectives: create educational content; collect feedback from learners; exchange complementary knowledge by resorting to external crowds; and by providing practical experience. Its most prevalent use concern educational material generation and assessment~\cite{alenezi,microlearning} and this is also true for CS-related curricula in higher education. For example, \cite{pirttinen2021exploring} describes a tool that can be used as a means to support teachers and students to create and review programming assignments. The main motivation behind such initiatives is premised on the potential benefits of crowdsourcing concerning optimizing the lecturing process and stimulating student involvement through knowledge co-creation and sharing, in line with contemporary learner-centered approaches to education~\cite{lambert1998students}.  

The current case study employs crowdsourcing in a project-based setting~\cite{project}, inviting students to grapple with a real-world problem --- that is creating a music knowledge base and a recommendation system. Tapping into the multiple prospective benefits of project-based learning, in \cite{khan2020completing}, the potential of resorting to crowdsourcing platforms for sourcing realistic tasks that can replace traditional assignments addressed to students of industrial design is discussed. The current case study aims to further investigate and leverage the employment of crowdsourcing in such a context, aiming at similar educational gains but from a different perspective: it approaches crowdsourcing not merely as a pool of possible pre-designed tasks, but rather as a methodology and technique that can be adapted to the specific course objectives and that is worth learning in its own right.


Applications in CS-related higher education curricula that adopt crowdsourcing as a means to provide practical experience in a setting relevant to the students' discipline are few. In \cite{kasumba2022application}, a crowdsourcing experiment conducted as part of a research project, involving data science students in rating homework reviews, had the unplanned effect of serving as a learning opportunity for students. 
In another case study presented in \cite{leinonen2020}, a system that employs crowdsourcing to enable students to create and review SQL assignments is introduced.
In \cite{chen2014quasi}, students of software engineering were assigned the task to test commercial software and through this process achieved industrial-strength training. The most common practice,  which is also followed by our work, is that instructors assume the role of the requester and students that of crowd-workers. An interesting exception to this is \cite{guo2018}, where graduate and undergraduate computer science students were asked to design and deploy their own crowdsourcing projects. The current case study adds to this line of work, by placing the focus on the challenges and possibilities of crowdsourced-enabled data enrichment in serving CS-relevant learning objectives and by contributing novel evidence and multi-dimensional insights grounded on an extensive analysis of feedback collected from students about multiple aspects.


Within the last years, there is an increasing number of initiatives that apply crowdsourcing in citizen science-oriented settings within formal and informal learning environments (schools and universities)~\cite{kloetzer2021learning}. Most such initiatives involve children and adolescents at the primary and secondary levels~\cite{roche2020citizen}, while citizen science projects in tertiary education remain fewer~\cite{vance2022citizen}. In an application of citizen science in an undergraduate environmental studies course~\cite{citizenhigher}, students were engaged in reporting roadkilled animals, thus gaining a deeper understanding of ecological problems and their solutions. Another case study~\cite{educsci12030151}, involving students from biology and environmental studies in field data collection, concludes that students enjoyed the learning process and improved their understanding of the domain as well as of crowdsourcing as a method for data collection. The crowdsourcing task selected for the current case study involves data enrichment of a music collection~\cite{gomes}.


In this respect, the current paper contributes to ongoing efforts~\cite{pmemodataset,mtg,Laurier,humphrey2018openmic,aljanaki2016studying}, many of which resort to crowdsourcing methods, to increase the availability and quality of annotated datasets that can be useful for prototyping systems for Music Information Retrieval (MIR) tasks~\cite{choi2016automatic} and particularly tasks concerning genre~\cite{costa2011music}, instrument~\cite{9051514}, and emotion recognition~\cite{liu2017cnn}. 

\section{Methodology for preparing the case study}

The case study was conducted as part of an assignment involving fourth-year undergraduate informatics students of the National Technical University of Athens who attended the course “Knowledge Systems and Technologies” in the spring semester of 2022. The main objective of the course is to introduce students to the fundamentals of description logic, methodologies for object-oriented knowledge representation, management, evolution, automated reasoning, and semantic data integration algorithms. Specific emphasis is given to the analysis of W3C standards for semantic data and knowledge representation (XML, RDF, OWL, etc), ontology engineering and applications of knowledge-based systems and intelligent web services~\cite{stamou2017ontological}. The course includes a semester-long multi-step assignment that aims to familiarize students with the above-mentioned concepts and associated tools via hands-on tasks. In line with these educational objectives, the case study set out to introduce concepts from digital CH as well as crowdsourcing to this purely CS-oriented curriculum and broaden the scope of the assignment towards an interdisciplinary direction. 

In its first step, the assignment focused on familiarizing students with the curated dataset and on engaging them as annotators in a crowdsourcing campaign as a means to enrich the dataset with additional useful knowledge. The main objective in this respect was for students to understand the shortcomings of real-world datasets and how raw, inadequate, or inconsistent forms of data can be transformed into well-structured, normalized, and inter-linked formats. Next, students were asked to transform the enriched data structure into a knowledge graph containing RDF triples and build an ontology that describes the data by constructing concepts, roles, axioms, and instances syntactically and semantically correct. Finally, students were solicited to use various methods to infer extra information and exploit it to make meaningful recommendations on music. 
In the following sections, we describe the incremental steps of the methodology we followed to set up the case study: the dataset curation (Section~\ref{subsec:dataset}); the definition of the enrichment tasks (Section~\ref{subsec:enrich}); and the organization of the crowdsourcing campaign (Section~\ref{subsec:crowdsourcing}). 
 
\subsection{Dataset curation}\label{subsec:dataset}

We decided to use the Europeana digital library~\footnote{https://www.europeana.eu/} to source the data that constituted the starting point of the crowdsourcing campaign and the subsequent assignment steps. Europeana currently aggregates more than 58 million records coming from CH Institutions (CHI) across Europe
CH items on the Europeana platform are described via a well-defined established metadata structure, the Europeana Data Model (EDM)~\cite{EDM}, which conveys important information about the items, such as their title, free text description, creator, etc. These metadata fields are essential for the accessibility and discoverability of the rich and disparate collections made available through the Europeana platform, helping users to find and understand the objects they are interested in. 

The first step towards the preparation of the case study concerned the curation of the dataset that would constitute the starting point of the crowdsourcing campaign and the subsequent assignment steps.
We started by scouting the music content available on the Europeana platform through the Europeana Search API, which provides a way to search for metadata records and media on the Europeana repository and supports advanced queries and filtering.

The following selection criteria were used to guide the curation process:
\begin{itemize}
    \item Quality of metadata that accompanied the music tracks. Metadata records on the Europeana platform often suffer from poor metadata.
    In order to build an initial knowledge base that can act as a
    sufficiently expressive starting point for further enrichment, we filtered out metadata records that lacked information considered essential for building an initial knowledge base/
    \item Quality and length of audio files. Audio files longer than 6 minutes were discarded, to filter out files that represented more than one music track 
    as well as to avoid assigning overly time-consuming tasks to students. The sound quality was also evaluated on sample files, which were considered indicative of the overall sound quality of a provider.
    \item Genre coverage. In order to serve the needs of the assignment and facilitate meaningful recommendations, the selection process aimed to cover a wide coverage of music genres (from classical and folk to rock and rap). 
\end{itemize}

By performing a series of API queries reflecting the criteria described above and evaluating a sample of the results, we ended up using data from the following CHIs: the "Internet Archive”, the "Internet Culturale / Biblioteca Nazionale Braidense - Milano”, and the “Fondazione Biblioteca Europea di Informazione e Cultura (BEIC)”. Eventually, 854 songs were collected. 
A post-filtering procedure on the curated metadata records was necessary since not all metadata fields included in the returned records are characterized by consistent values. 
The post-filtering process resulted in the metadata properties shown in Table~\ref{tab:metadata}. 


\begin{table*}[!t]
\tiny
\centering
\caption{Metadata specifications}
\begin{tabular}{m{0.1\columnwidth}|m{0.27\columnwidth}|m{0.42\columnwidth}}
\hline
\rowcolor [gray]{0.7}Property name & Correspondence to EDM property & Specification  \\
\hline\hline
EuropeanaID & rdf:about & the music track’s unique Europeana record ID\\
\hline
Title & dc:title &the music track's title\\
\hline
Year & dc:date &the year when the performance was recorded\\
\hline
Duration & ebucore:duration &the duration of the track in milliseconds\\
\hline
Composer & dc:creator &the composer of the music track\\
\hline
DateOfBirth & rdaGr2:dateOfBirth & the date of birth of the music track's composer\\
\hline
DateOfDeath & rdaGr2:dateOfDeath & the date of death of the music track's composer\\
\hline
Biography & rdaGr2:biographicalInformation & the biography of the music track's composer\\
\hline
Publisher & dc:publisher & the publisher of the music track
\\
\hline
Place & skos:prefLabel & the place where the performance was recorded\\
\hline
\end{tabular}
\label{tab:metadata}
\end{table*}


\subsection{Definition of the enrichment goals}\label{subsec:enrich}

The information conveyed by the original metadata properties sourced from the Europeana platform is quite limited, allowing only for quite basic queries and restricting the potential for their meaningful further exploitation. In order to enable higher flexibility, richer ontology structures, and more reliable recommendations, the curated dataset has to be enriched with more information that can be exploited by the later stages of the assignment (see Section~\ref{sec:knowledgebase}). At the same time, the more extensive and specialized information is added, the more expert knowledge, effort, and time is required. For example, retrieving detailed information about the performance and featured artists (e.g. singers, musicians) requires dedicated research. It should also be noted that performing raw audio analysis for extracting sonic characteristics, such as "instrumentalness" or "danceability", used by established music recommendation systems~\cite{spotify} is beyond the scope of the specific CS course. Similarly, taking into consideration the size of the class and time constraints, the analysis of user taste profiles as a means to inform recommendations on music was not considered as part of the assignment.

Weighing in the above considerations and in order to achieve a middleground between desired richness and feasibility, the manual enrichment process focused on collecting data along the following three aspects: "Emotion", "Genre" and "Instruments". These enrichment goals were formulated as crowdsourcing tasks to be carried out by students via their participation in an appropriately designed campaign (see Section~\ref{subsec:crowdsourcing}). The terms for all metadata fields  and respective type of tasks correspond to Wikidata URIs and were selected based on specific criteria, as explained below. 

\begin{figure}[!b]
    \centering    \includegraphics[scale=0.35]{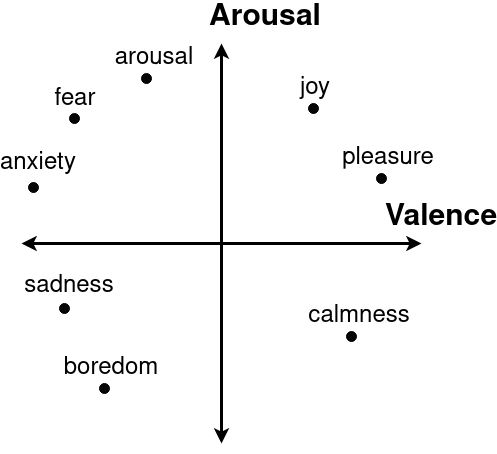}    \caption{Emotion tags in circumplex model}
    \label{fig:sysarch}
\end{figure}

Emotion reflects how the audience feels when listening to a music track and can be exploited for making meaningful music recommendations. Obviously, this is partly a subjective issue - every person perceives a music piece in their own way, although a majority of people would usually agree whether a song is melancholic or joyful.
The subjective dimension of emotion is an additional reason why an aggregated opinion by the crowd can help us derive an "average" metric about what kind of emotion a song gives rise to.
In order to represent emotion within music, we based on the circumplex model developed by James Russell \cite{russell}. The model is oriented around two dimensions: arousal represents the vertical axis and valence represents the horizontal axis. The emotion values-tags that we used included: \textit{Arousal, Joy, Pleasure, Calmness, Boredom, Sadness, Anxiety and Fear}. Their place on Russell's model is shown in Fig.~\ref{fig:sysarch}.

Genre is a characteristic that is commonly used to organize music tracks and is exploited by music recommendation systems, sometimes in combination with emotion~\cite{1706489}. 
Taking into account the fact that our annotators are not music experts, we selected tags that represent typical music categories  (e.g. \textit{Rock}) instead of tags with highly specific usage (e.g. \textit{Alternative Rock}). Considering also the coverage of the curated dataset, the following controlled list of terms was used: \textit{Pop, Rock, Country, Classical, Opera, Instrumental, Funk, Hip-hop, Reggae, Jazz}, and \textit{Traditional Folk}.             

The instruments used in a track are another important musical characteristic. Compared to the other categories mentioned above, it is the characteristic that requires the most familiarity with music. For the musical instrument annotation task, we used 12 different terms. Similarly to our approach on representing genre, we selected rather high-level terms with a broad application (e.g. \textit{Brass} instead of \textit{Trumpet, Trombone etc}) as well as basic instruments (such as~\textit{Piano}), whose sound can be adequately recognized even by non-experts. Considering also the most common instruments appearing in the curated dataset, we ended up with the following instrument tags: \textit{Piano, Electric Guitar, Acoustic Guitar, Drums, Synthesizer, Violin, Harmonica, Banjo, Bass, Woodwind, Brass, Orchestra}. 


In addition to the three musical characteristics represented via controlled lists of Wikidata terms as mentioned above, students were also given the possibility to add free text comments about the music track they listened to. These comments could be exploited in later stages of the assignment, complementary to the "Emotion" tag, in order to further enrich the dataset using Natural Language Processing (NLP) techniques for sentiment analysis (see Section~\ref{sec:knowledgebase}). 

 \subsection{Setup of the campaign by using the CrowdHeritage platform}
 \label{subsec:crowdsourcing}

After preparing the data and defining the enrichment objectives, the next step was to set up and run the crowdsourcing campaign that would allow students to perform the actual enrichment tasks in a well-defined and collaborative way. More specifically, students were invited to listen to the music tracks, recognize their musical characteristics, and tag the items appropriately by selecting values from the controlled lists of terms defined above. The open-source CrowdHeritage platform~\cite{citizenhe} was used to this end. The platform  supports the organization of online crowdsourcing campaigns for the enrichment and validation of CH metadata. Through a user-friendly interface that supports playful features such as leaderboards and rewards, users are invited to add new annotations or validate (via crowd-voting) existing ones produced either automatically by AI tools or added by other users of the platform.  This validation input is further analyzed to identify questionable annotations and users with malicious or unreliable behavior, with crowd-voting  acting as a means of peer-reviewing. 

A campaign with concrete instructions was set up, which run for 18 days. The campaign setup included the import of the curated dataset; the definition of the annotation tasks by making use of the vocabularies/lists of controlled terms as defined in Section~\ref{subsec:enrich}; and the specification of the campaign's overall objective, associated instructions, duration etc.
The curated dataset was divided into eight sets of items with respective micro-tasks, in order to ensure balanced contributions by participants across the data. Students were advised to semantically annotate about 80 music tracks each and were encouraged to also add comments expressing additional information and their thoughts in free text. The completed campaign can be accessed here: \href{https://crowdheritage.eu/en/music-citizen}{Campaign}.

\section{Using the enriched dataset to build and query a music knowledge base}\label{sec:knowledgebase}

At first, the annotations collected from the campaign underwent a review and filtering procedure (see Section~\ref{subsec:enrich_results}) and were then parsed and embedded as new properties to the EDM metadata records. The resulting enriched dataset was moderated (see Section~\ref{subsec:enrich_results}) and provided to the students as a CSV file. Students were advised to transform the tabular data to a knowledge graph~\cite{Hogan_2022}. 
The next step was to build an ontology linked with the graph using the Protégé editor~\cite{protege}. The objective of this step was to teach the students how to structure the conceptual knowledge that can be inferred from the individual track instances into a generalized semantic model (as captured by the ontology) expressed in the form of concepts and properties. For example, the concepts "Song" and "Composer" can be used to represent the set of all items-tracks and composers respectively, while the property "hasComposer" can be used to connect a song with its composer.

The transformation of the dataset into a graph associated with an accompanying ontology opened the possibility for further automatic enrichment of the data using semantic techniques and enabled the support for advanced queries. The main techniques for further automatic data enrichment introduced to the students included: (i) accessing additional knowledge from external Linked Open Data resources; (ii) applying NLP on the free text comments; and (iii) extending the ontology by creating new concepts through axioms. Regarding (i), the students were advised to exploit the Wikidata URIs included in the metadata records and use the Wikidata SPARQL endpoint~\footnote{https://query.wikidata.org/bigdata/namespace/wdq/sparql} 
in order to retrieve additional information and link it to the knowledge graph's entities. For example, using the composer's name, the students could construct a SPARQL query that fetches the artistic movements that characterize this composer or the location where the composer was born. Regarding (ii), students were solicited to apply a sentiment intensity analysis model~\cite{vader} to analyze the free text comments added by students through the campaign and extract additional sentiment metadata features. 

This additionally retrieved information was incorporated into the knowledge graph and used as an extra characteristic 
for identifying tracks that may be relevant for the user. As for (iii), the 
students were instructed to create novel concepts in the ontology, in order to support more expressive queries by combining existing information. For example, the concept \textit{CalmJazzSong} can be defined via an appropriate axiom that groups together music tracks that have \textit{Jazz} as its genre and \textit{Calmness} as a relevant emotion; while the concept \textit{NineteenthCenturyComposer} can be used for representing composers who were born in the nineteenth century.  
At the point that the students had created a music knowledge base by linking their extended ontology with the enriched knowledge graph, they were able to apply SPARQL queries to it with the aim to identify tracks similar to a given track based on multiple criteria 
and thus make recommendations.

\section{Results and Evaluation}
 
Overall, the crowdsourcing campaign involved 98 participants, 68 males, and 30 females, all of whom were students of the course “Knowledge Systems and Technologies” of age 21-23 years old. Below, we provide an overview of the annotations contributed during the campaign. We then discuss the information collected via an online survey that was conducted after the completion of the campaign.

\subsection{Campaign results}\label{subsec:enrich_results}

The campaign led to the addition of 8399 annotation tags in total, while there have been 49351 up-votes and 495 down-votes of annotations added by other users. A moderation process was necessary to review and filter out the results which were considered of questionable validity.
The number of up- versus down-votes received by an annotation was used as the main criterion to assess its reliability and resolve issues of ambiguity, subjectivity, malicious or irresponsible behavior via a majority vote. The annotations' moderation took place by making use of the validation editor provided by the CrowdHeritage platform (Section~\ref{subsec:crowdsourcing}), which allows campaign organizers to review the annotations produced during a campaign and filter them according to their own acceptance criteria.  
During the moderation process, only the two top-ranked annotations per \textbf{Emotion} and \textbf{Genre} were kept and only if these had an up- versus down-votes difference of at least two. For the \textbf{Instruments} property in particular, only values with a votes difference above five were kept. This rather strict pruning criterion was decided because many students mentioned in their feedback (see Section~\ref{subsec:eval}) that they did not have the necessary expertise to recognize musical instruments. In addition to this filtering process, the annotations of a random sample of 80 music tracks were reviewed by two music experts, who concluded that the enriched metadata were of high quality. 

\begin{figure}[!t]
    \centering
    \includegraphics[scale=0.4]{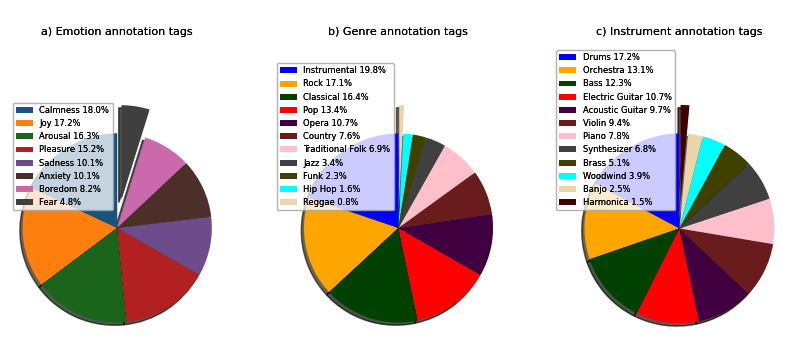}
    \caption{Annotation tags per category from the campaign}
    \label{fig:charts}
\end{figure}

As a result of the post-filtering process, 5147 annotations were kept: 1248 of them refer to genre tags, 1643 to emotion tags, 1422 to instrument tags and 834 represent free text comments. In Fig.~\ref{fig:charts} the statistics of annotation tags per metadata property are presented. For the instrument annotation task, we notice that tracks with knowable and distinguishable sounds such as \textit{Drums} and \textit{Orchestra} are the most annotated ones. Furthermore, the distribution of the Genre tags demonstrates that \textit{Instrumental, Rock, Classical} and \textit{Pop} are the most dominant tags. As for the emotion property, we observe that positive emotions are the most common, a finding that confirms the bias towards positive emotions in music datasets discussed in previous work~\cite{Cano}.

In order to further assess the quality of the collected metadata, we analyzed them by using association rules. We applied the Apriori algorithm~\cite{agrawal1994fast} on the total of the tag element sets in order to study their correlation. We took into account the support metric in order to observe the most frequently observed pair tags. The support metric is calculated by counting the occurrences of both tags appearing in the same set and dividing them by the total number of set. Table~\ref{tab:association} shows which paired tags appear more frequently. We observe that the most common paired tags reflect intuitive knowledge about music (e.g. Rock-Drums, Rock-Electric Guitar, Classical-Orchestra), while paired tags connecting emotion and genre (e.g. Calmness-Instrumental) are also in accordance with prior findings~\cite{worlu2017predicting}.


\subsection{Open annotated dataset}

The annotated dataset that resulted from the crowdsourcing campaign and the respective post-filtering process 
can be valuable for the prototyping and evaluation of MIR systems.
To this end, the collected music tracks, metadata, and moderated enrichments are made openly available, so that they can be freely reused as data amenable for computational purposes. All three categories of enrichments-annotations (genre, instruments, emotion), as well as the properties retrieved from the EDM (see Section~\ref{subsec:dataset}), are included in a single dataset. All  metadata with their enrichments are made available under a CCO license. We provide the GitHub repository for the dataset: \href{https://github.com/vaslyb/MusicCrowd}{Dataset}.

All tracks are annotated with respect to genre, emotion, and identified instruments using the value lists described in Section~\ref{subsec:dataset}. It should be noted that the rather strict filtering criteria already mentioned ensure that only annotations for which there is high certainty for their validity are maintained. The filtering based on the up-/down-voting of annotations, in particular, compensates for factors commonly identified as leading to poor annotations, such as inattentive labeling, listener fatigue, or other errors~\cite{soleymani20131000}. 

\subsection{Evaluation by participants}\label{subsec:eval}

The online survey addressed to students consisted of a combination of closed and open questions. First, we aimed to understand how the students experienced the crowdsourcing process as a part of their mini-project assignment. Relevant questions investigated: the degree to which the crowdsourcing objectives were lucid; what students identified as the main benefits of introducing a crowdsourcing campaign in the assignment; the degree and ways in which the process improved or extended students' knowledge and skills; the kind of feelings their participation gave rise to; and the types of difficulties they experienced when performing their tasks. Secondly, we aimed to collect feedback about the CrowdHeritage platform as a tool for contributing to crowdsourcing campaigns. Questions in this track focused on the overall usability of the platform; the usefulness and efficiency of different sub-components/functionalities; and on identifying certain shortcomings and collecting recommendations for further improvements. 

35 students provided answers to the online questionnaire (5 females and 30 males). This low participation in the survey in comparison with the number of students who contributed to the campaign (36\% of the campaign participants) is mainly attributed to the fact that answering the questionnaire was not seen as an integral/necessary step of the course assignment. It should be noted, however, that many students opted to use the free commenting functionality of the CrowdHeritage platform as a means to express their perceptions and provide feedback. 

\begin{figure}[t]
  \begin{minipage}{0.5\linewidth}
    \centering
    \includegraphics[scale=0.4]{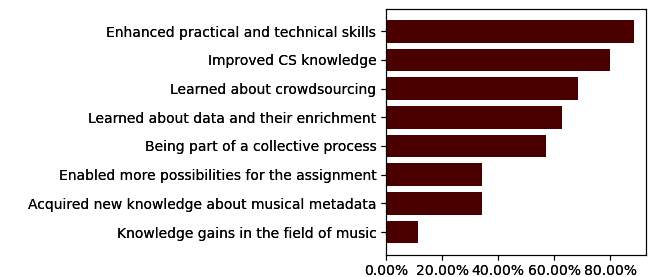}
    \caption{Benefits gained by students}
    \label{fig:hist}
  \end{minipage}
  \hfill
  \begin{minipage}{0.43\linewidth}
    \centering
    \tiny
    \begin{tabular}{p{0.15\linewidth}p{0.5\linewidth}}
    \rowcolor [gray]{0.7}Support & Pair of Tags\\
    \hline
    0.201 & Joy, Drums\\
    \hline
    0.192 & Rock, Drums\\
    \hline
    0.185 & Drums, Arousal\\
    \hline
    0.181 &	Electric Guitar, Drums\\
    \hline
    0.162 &	Electric Guitar, Rock\\
    \hline
    0.147 &	Classical, Instrumental\\
    \hline
    0.147 &	Classical, Orchestra\\
    \hline
    0.145 & Calmness, Instrumental\\
    \hline
    0.144 & Joy, Arousal\\
    \hline
    0.138 &	Bass, Drums\\
    \hline
    0.134 & Instrumental, Orchestra\\
    \hline
    \end{tabular}
    \captionof{table}{Frequent paired tags}
    \label{tab:association} 
  \end{minipage}
\end{figure}

The objective of the campaign as well as of the overall assignment was well-understood by the students (97\% described the objectives as "very clear/clear" and 3\% as "clear enough"). 52\% of the students described their participation experience as interesting or very interesting, 37\% as neutral (neither boring nor very interesting) and 11\% as boring. All students expressed that they had some knowledge gains: 77\% of the students declared that their knowledge and skills were improved and expanded to a very large or large degree and 33\% to a sufficient degree. 88\% of the students stated that they enhanced their practical and technical skills, e.g. learned how to use certain technological frameworks such as writing in Python to manipulate data, and 80\% that they improved their CS scientific knowledge, e.g. with respect to semantic web principles (see Fig.~\ref{fig:hist}). Highly appreciated benefits also included: learning about the potential of crowdsourcing and how this is conducted (69\%);  gaining a deeper understanding of the data, their shortcomings, and the value of their enrichment (63\%); the participatory elements that crowdsourcing added to the assignment (56\%); and facilitating, through the data enrichment, more interesting things in later stages of the assignment (34\%). Only 34\% stated that they acquired new knowledge about cultural and musical metadata (e.g. their structure, properties). Knowledge gains in the field of music (e.g. learning about new songs, genres, to identify instruments) were mentioned only by 11\% of the students. All students declared that they would consider employing crowdsourcing as a means for data enrichment in the future.


The most commonly mentioned factor that hampered students' degree of engagement concerned the fact that certain tasks, and particularly the identification of genres and instruments, required a degree of music sophistication which many students did not possess (46\% of the students encountered this difficulty). This finding is aligned with the observations made in previous work discussing music content annotation campaigns~\cite{samiotis}, which express the concern that crowd workers are often expected to annotate complicated music artefacts that demand certain skills that participants may lack. Moreover, the completion of all the annotations tasks expected by a user was perceived as too time-consuming (35\% of students pointed to this issue). Some students mentioned that, in some cases, the available choices from the controlled vocabulary lists were not sufficient to convey what they would like to express, while others identified some music tracks as being of poor quality.

Besides the questionnaire, the free text commenting functionality supported by the CrowdHeritage platform can also be seen as an indicator of the engagement of the participants. Overall, students added 834 comments under different tasks-tracks. The high number of free comments demonstrates that students felt the need to express their thoughts besides the framework of the strictly defined enrichment tasks and thus reflects a genuine sense of involvement. 



\section{Conclusions and discussion}

The current work exemplifies crowdsourcing as a promising practice in CS-related curricula of higher education, illustrating how it can be embedded in a homework mini-project. 
The methodology followed, the tools used, and the overall experience accrued can pave the way for embracing crowdsourcing in other frameworks within the scope of CS curricula. For example, crowdsourcing could be used to in combination with ML tasks, to enrich an ontology and its relations, or in a course on human-computer interaction, with an emphasis on the UX features that should characterize platforms used for conducting crowdsourcing tasks. The insights and recommendations for improving the CrowdHeritage platform collected by students in the current case study already point to interesting ideas towards this direction. Parts of the methodology that was followed can also be useful for educational purposes in other disciplines, besides CS, such as digital humanities. An interesting direction that can be explored in various disciplines is ways to engage students as requesters in the  preparation phases of the crowdsourcing lifecycle and ask them to design their own crowdsourcing projects in order to solve a specific problem. 

Revisiting the two main strands of inquiry we set out to investigate, concerning the educational and the engagement implications of crowdsourcing, the assessment of the results and the feedback received from the students point us to some interesting conclusions. Transparency and clarity about the objectives of the crowdsourcing process and its functioning in the overall flow of the CS assignment was considered crucial by the students so as to understand the relevance of the project and how their contributions would be used and their skills would improve, thus attaining their interest and investment in the project. 

The knowledge gains from the crowdsourcing enrichment process are evidenced by the deep understanding which students acquired about the metadata structure and its characteristics as well as the gradual process they followed to construct a knowledge graph and an ontology of increasing richness and expressiveness. The multiple and genuine ways in which students exploited the enriched data to develop complex concepts and queries and build added-value features also attest to the conclusion that the assignment served its educational purpose. As manifested by the students' responses, what was mostly appreciated concerned competences which advanced their CS expertise. Students also got acquainted with the practical technical challenges behind crowdsourcing, especially concerning the UX features that make a platform successful. This is reflected in the apt feedback and recommendations students provided about the CrowdHeritage platform.
Knowledge benefits from performing the music annotation tasks themselves, in their role as crowd workers, were less acknowledged. 

Concerning the engagement dimension, feelings appeared to be mixed. Although almost all students liked the incorporation of the crowdsourcing campaign in the assignment and perceived the platform as user-friendly, almost half of them described their participation experience as neutral or even boring. Crowdsourcing was mostly appreciated in a rather instrumental way, as a practical means to achieve an interesting end. This can be partly attributed to the quite demanding goal that was set and the fact that many students felt that certain tasks required quite advanced music sophistication that they lacked. 
Even so, we cannot ignore the fact that the commonly praised participatory and affective benefits of citizen science and crowdsourcing were not the most cherished ones by students. This resonates with recent criticisms on the way in which crowdsourced citizen science is touted as an enjoyable and participatory experience, while at the same time its labor ramifications and the repetitive or mundane nature of the crowdsourced tasks are understated~\cite{kidd2018public,del2016crowdsourcing}. Further experiments and more in-depth evaluation in  higher education settings is required to shed more light on this aspect. 

Although the current study lays its primary focus on the role and impact of crowdsourcing within the CS higher education community, the publication of the carefully filtered annotated dataset is also an important outcome that can prove helpful for the research and ML communities. An inspection of the annotations' characteristics allowed us to draw some useful insights concerning the human subjects' behavior, the correlation between tags, and the overall annotations' quality. Further work is required in order to yield the dataset readily amenable for the development and evaluation of MIR models. An expansion of the dataset would enhance and widen its usefulness for computational models. Further data reliability analysis and experimentation is required to demonstrate the dataset's validity and possible usages and to establish a benchmark for the MIR community.

\section*{Acknowledgments}
The work is co-funded by the Erasmus+ Program of the European Union, project number 2020-1-BE02-KA203-074727. We thank the students of the course “Knowledge Systems and Technologies” for their contributions to the crowdsourcing campaign and valuable feedback.
\bibliography{sample-base}

\end{document}